\documentstyle[aps,amssymb,amsmath,preprint,graphicx]{revtex}

\begin{document}
    
\draft

\title{ Velocity Correlations, Diffusion and Stochasticity 
in a One-Dimensional 
System}
\author{V. Balakrishnan\footnote{Permanent address: Department of Physics, 
Indian Institute of Technology-Madras, Chennai 600 036, India}}
\address{Centre for Nonlinear Phenomena and Complex Systems, 
Universite Libre de Bruxelles, C.P. 231, Bd. du Triomphe, 1050 
Brussels, Belgium}
\author{I. Bena and C. Van den Broeck }
\address{ Limburgs Universitair Centrum, 
B-3590 Diepenbeek, Belgium}

\date{\today}
\maketitle

\begin{abstract}
We consider the motion of a test particle in a one-dimensional system 
of equal-mass point particles. The test particle plays the role of a 
microscopic ``piston" 
that separates two hard-point gases with different concentrations and  
arbitrary initial velocity distributions. In the homogeneous case when 
the gases 
on either side of the piston are in the same macroscopic state, we 
compute and analyze the stationary velocity autocorrelation function $C(t)$.
Explicit expressions are obtained for certain
typical velocity distributions, serving to elucidate in particular the 
asymptotic behavior of $C(t)$. It is shown that the occurrence 
of a non-vanishing probability mass at zero velocity is necessary for
the occurrence of a long-time tail in $C(t)$. The conditions under 
which this is a $t^{-3}$ tail are determined. Turning to the 
inhomogeneous system with different macroscopic states on either side
of the piston, we determine its effective diffusion coefficient from 
the asymptotic behavior of the variance of its position, as well as the
leading behavior of the other moments about the mean. Finally, we 
present an interpretation of the effective noise arising from the 
dynamics of the two gases, and thence that of the stochastic process to 
which the position of any particle in the system reduces in the thermodynamic 
limit.\\
\end{abstract}


\section{Introduction}

\par  Exactly solvable ``toy" models are important in understanding the 
dynamic behavior of complex systems made up of a large number of particles.
They also allow us to establish and understand the limitations of the 
approximations used in general to deal with  
systems of interacting particles, such as the Boltzmann equation.
One such model, consisting of identical hard-point particles 
moving on a line and interacting through elastic collisions, was 
introduced several decades ago \cite{model}.
Based on the  observation that the particles merely exchange velocities
in a collision, 
Jepsen \cite{jep} was able
to calculate explicitly several properties of a gas of such particles. 
Subsequently, Lebowitz and co-workers \cite{lebo1}- 
\cite{lebo3} refined and extended these calculations to include, among other 
aspects, a  comparison with the results of the Boltzmann approximation.
Very recently, this model system has been revisited by
Piasecki and Gruber \cite{piagru},\cite{pia}, their main motivation 
being the
construction of a one-dimensional
analog of the  ``adiabatic piston"
\cite{adpiston}, with a central  particle playing the role of a piston 
separating gases at different temperatures and densities to its left and 
right. Attention  has also been 
focused on several related models from the point of 
view of the applicability of the Fourier law for heat flux in 
one-dimensional systems \cite{fourierlaw}. Although the main 
interest in these specific contexts is in the case of an arbitrary ratio of the 
masses of the piston and a gas particle, the equal-mass case being a 
singular one in some sense, the analytical tractability of the latter makes it a 
valuable ``theoretical laboratory" from which much can be learnt. In 
particular, it offers a model in which (i) one can pass to the 
thermodynamic limit in a rigorous manner, and (ii) 
the effect of recollisions can 
be {\it completely and exactly} taken into account. 

The renewed interest in
the model prompts us to revisit it and extend its analysis. 
We focus on the 
motion of the central particle (which we shall refer to as the ``piston" 
for brevity). A brief recapitulation of the relevant results
from earlier work is given in Section II. In Section III we
compute and study the stationary velocity autocorrelation function 
$C(t)$ of the piston in  the homogeneous case, when 
the gases 
on either side of the piston are in the same macroscopic state (i.e., 
equal densities and arbitrary but identical velocity distributions),
to amend and extend earlier results. 
The explicit expressions obtained for certain archetypical velocity 
distributions  help us analyze the asymptotic behavior of $C(t)$ to 
determine exactly when a power-law decay may be expected, and when the 
latter is a $t^{-3}$ tail. Turning to the 
inhomogeneous system with different macroscopic states on either side
of the piston, in Section IV we determine its effective diffusion coefficient 
from the asymptotic behavior of the variance of its position, as well as the
leading behavior of the other moments about the mean. Finally, in 
Section V we show that there is an appealing and direct 
interpretation of the effective noise to which the many-particle 
interactions reduce in the thermodynamic limit as far as one-particle 
dynamics is concerned, and thus of the stochastic process 
represented by the position of any particle in the system. 

\section{Recapitulation of earlier work}
 
For ready reference, we record briefly the relevant results 
of earlier work that are needed for what follows, 
using the convenient notation employed 
by Piasecki \cite{pia}.
One starts with a system of a finite
number of particles, located in the
interval $[-L,\,L]$ of the $x$-axis, with free boundary conditions. 
The piston is initially located (without loss of generality) at 
$X_{0}=0$ and  
has an arbitrary initial velocity 
$V_{0}$.  
The $N^{-}$ particles to its left and ${N}^{+}$ particles to its 
right are initially at uniformly distributed random positions $X_j\,
(j=-N^-, \ldots, 0,
\ldots, N^+)$,
with independent, 
identically distributed velocities $V_j$ drawn 
from  normalized velocity distributions $\phi^{\pm}(V)$ respectively.  
As the
particles merely exchange velocities upon collision, at any instant of time 
the piston is on one of the ``free" trajectories $X_j + V_jt$. This, 
together with the fact that the particles cannot move across each 
other, suffices to solve \cite{jep} the problem of determining, among other 
quantities, the phase space 
distribution function of the piston (or that of any of the other particles) at 
any time $t > 0$ by averaging over the initial positions and 
velocities of the gas particles on both sides of the piston 
\cite{jep},\cite{lebo1}.
In the thermodynamic limit $L \rightarrow \infty$ and  
$N^{\pm}\rightarrow \infty$ such that one has finite densities
$\lim N^{\pm}/L\,= n^{\pm}$ on the left and right of the piston,
the conditional one-particle distribution function  
of the piston is found to be
\begin{eqnarray}
&&\hspace{-1cm} P(X,V,t|0,V_0,\,0)= \nonumber\\
&&e^{-t[n^-\alpha (V_0) + n^+ \beta
(V_0)]} I_0 \left(2t [n^-
\alpha (V_0) n^+ \beta (V_0)]^{1/2} \right) \,\delta (X-V_0t) 
\,\delta(V-V_0)\nonumber\\ 
&+& \,e^{-t[n^-\alpha
(X/t) + n^+\beta (X/t)]} \left\{ [n^- \phi^-
(V) \theta (Vt - X) \theta
(X-V_0t) \right. \nonumber\\ 
&+&  \left. n^+
\phi^+ (V) \theta (X-Vt) \theta (V_0 t-X)]  \right. \, I_0 
\left(2t
[n^- \alpha (X/t) n^+ \beta (X/t)]^{1/2} \right) \nonumber\\
&+&  \left[  n^- \phi^- (V)
\theta (Vt-X)
\theta (V_0t - X)
\left(n^+ \beta (X/t)/n^-\alpha (X/t)\right)^{1/2} +\,n^+ \phi^+
(V) \theta (X-Vt) \,\right.\nonumber\\
&& \times \left.
\theta (X-V_0t) 
\,\left(n^- \alpha (X/t)/n^+
\beta (X/t)\right)^{1/2} \,\,\right] \,\left. I_1
\left(2t [n^- \alpha (X/t) n^+ \beta (X/t)]^{1/2}\right) \right \} 
\,\,,\nonumber\\
\label{4}
\end{eqnarray}
where
\begin{equation}
\alpha (W) = \int\limits_W^\infty dU \phi^- (U) 
(U-W)\,\,\,\,, \,\,\,\,\, \beta(W) = \int
\limits_{-\infty}^W dU \phi^+ (U)(W-U) \,\,,
\label{5}
\end{equation}
and $I_{n}$ is the modified Bessel function of order $n$.
For  $V_0= 0$,  this is the result recorded in  Ref. \cite{pia}. 

The velocity distribution of the piston is obtained by integrating 
$P(X,V,t|0, V_0,\,0)$ over $X$, and is given by
\begin{eqnarray}
& & \hspace{-0.5cm}P(V,t|0, V_0,\,0) = 
e^{-t\left[n^-\alpha(V_0)+n^+\beta(V_0)\right]} I_0\left(2t [n^-
\alpha (V_0) n^+ \beta (V_0)]^{1/2} \right) \, \delta (V-V_0) \,\nonumber\\
& & +\,\,t \int_{-\infty}^{\infty} dW\,e^{-t\left[n^-\alpha (W) + n^+\beta (W)\right]} 
\left\{ \left[n^- \phi^- (V)
\theta (V-W) \theta (W-V_0) \right. \right. \nonumber\\
&& +\, \left. n^+ \phi^+ (V) \theta (W-V)
\theta(V_0-W)\right]  \, I_0\left(2t [n^-
\alpha (W) n^+ \beta (W)]^{1/2} \right)   
\nonumber\\ 
&&+ \left[n^- \phi^-(V) \theta
(V-W) \theta (V_0-W) \left(n^+ \beta(W)/n^-\alpha(W)\right)^{1/2} 
+ \,n^+ \phi^+(V) \theta (W-V) \right. 
\nonumber\\
&& \left. \left. \times \,\theta (W-V_0) \left(n^-
\alpha(W)/n^+ \beta(W)\right)^{1/2}\,\,\right] \,I_1 \left(2t [n^-
\alpha (W) n^+ \beta (W)]^{1/2} \right) \right \} \,\,.
\label{6}
\end{eqnarray}
The stationary velocity distribution to which this tends as 
$t \rightarrow \infty$ is found \cite{pia} by using the leading term in the 
asymptotic expansion 
of $I_{n}(z)$ for large argument, and carrying out the integral 
over the resulting Gaussian 
peaked at 
$W = \overline{W}$, where $\overline{W}$ is the (unique) root of the equation
\begin{equation}
n^- \alpha(\overline{W}) = n^+ \beta (\overline{W})\,\,.
\label{10}
\end{equation} 
The  
normalized asymptotic velocity distribution of the piston is then 
found to be 
 \begin{equation}
P_{\infty} (V) = \frac{n^- \phi^-(V) \theta(V-\overline{W}) + n^+ \phi^+(V)
\theta (\overline{W}-V)}{\Xi(\overline{W})} \,\,,\label{7}
\end{equation}
where 
\begin{equation}
\Xi (\overline{W}) = -n^-\alpha'(\overline{W}) + n^+\beta'
(\overline{W}) = n^-|\alpha' (\overline{W}) | +n^+ \beta' (\overline{W})
\label{8}
\end{equation} 
is the normalization factor. In general, therefore, $P_{\infty} (V)$ 
has a finite discontinuity at $V = {\overline{W}}$. The
asymptotic drift velocity of the 
piston, defined as
$\langle V({\infty}) \rangle = \int_{-\infty}^{\infty} dV\, 
V\,P_{\infty}(V)\,$, 
is then trivially seen to be equal to  $\overline{W}$ itself, on 
re-writing the definition of the latter in Eq. (\ref{10})
as 
\begin{equation}
\overline{W} = \frac{n^- \int\limits_{\overline{W}}^\infty dU U \phi^-(U) +
n^+ \int
\limits_{-\infty}^{\overline{W}} dU U \phi^+ (U)}{n^- \int
\limits_{\overline{W}}^\infty dU \phi^-(U) +
n^+ \int \limits_{-\infty}^{\overline{W}} dU \phi^+(U)} \,\,.\label{11}
\end{equation} 
In particular, for Maxwellian distributions $\phi^\pm (V)$ at
temperatures $T^\pm$,
one has $\overline{W} \gtrless  0$ according
as $n^{-}\sqrt{T^-} \gtrless n^{+}\sqrt{T^+}$.
Since $n (k_{B}T^{\pm})^{1/2}$ 
is essentially the linear friction coefficient
of the corresponding gas, it is
this {\it dynamic} property, rather than the pressure, that
determines the direction of the drift of the
central particle. The equality $n^{-}\sqrt{T^-} = n^{+}\sqrt{T^+}$ 
as the condition necessary for the piston to have a drift-free,
purely diffusive motion asymptotically, has already been pointed out 
by Jepsen \cite{jep}.

We can proceed to show that, in the limit $t \rightarrow \infty$, 
{\it every} particle attains the new stationary velocity distribution 
$P_{\infty}(V)$ (in the thermodynamic limit, of course). This remains 
true even if $\phi^{-} = \phi^{+} = \phi$, but $n^{-} \neq n^{+}$. 
Thus, although the system under study does not have strong mixing 
properties, in the sense that the set of ``free" trajectories $\{X_{j} + 
V_{j}t\}$ is fixed for all time once the initial values $\{X_{j}\}$ and $\{V_{j}\}$ 
are specified, the effect of the collisions of the particles with the 
piston and among themselves is to mix
the initial velocity distributions on the right and left of the 
piston. This happens by a ``diffusion" of the  ``interaction front"
of the piston with the other particles, through successive  
collisions. 
Only in the special case in which
\begin{equation}
n^- = n^+ = n,\ \phi^- = \phi^+  = \phi \,,
\label{symm}
\end{equation}
i.e., when  the 
gases on both sides of the piston are initially in the same macroscopic state,  
does $P_{\infty}(V)$ reduce to $\phi(V)$ itself.
We refer to this as the ``homogeneous system" in 
what follows, in contrast to the more general inhomogeneous case.
For simplicity, we 
shall also assume that $\phi(V) = \phi(-V)$ throughout this paper.

\section{Velocity autocorrelation in the homogeneous system}

We now study the {\it stationary} velocity autocorrelation function 
of the piston, which 
requires that the initial distribution be a stationary one. As 
mentioned above, this only 
happens in the homogeneous case, to which we therefore
restrict ourselves in the rest of this section. The results to be
obtained are in fact valid for {\it any} particle in the 
system.

Since $\overline{W} = 0$
in this case, the autocorrelation 
function is given by 
\begin{equation}
C(t) = \langle V(t_{0})
V(t+t_{0})\rangle  =\int_{-\infty}^{\infty} dV \int_{-\infty}^{\infty} dV_{0} 
\,\,V_{0}\, V 
\,P(V,\,t|0, V_{0},\,0) \phi(V_{0})\,\,.
\label{correlation}
\end{equation}
This quantity has been considered in earlier work both for a Maxwellian
$\phi(V)$ \cite{jep} and in more general terms \cite{lebo1}, and 
been found to exhibit a $t^{-3}$  power-law tail. However, the 
conclusions regarding the conditions under 
which this happens require modification; nor is the exponent of the 
power law invariably equal to $-3$. We therefore 
analyze the question afresh, extending and amending some of these 
earlier results. We also give a
physical interpretation of the circumstances leading to a power 
law decay of  $C(t)$. 

On inserting Eq. (\ref{6})  in Eq. (\ref{correlation}) 
and simplifying, 
we obtain the expression
 \begin{eqnarray}
C(t) &=&\, \int_{-\infty}^{\infty} dW 
\,e^{-nt \left[\alpha (W) + \beta (W)\right]} \,
\left\{ I_0 \left(2nt
[\alpha(W)\beta(W)]^{1/2}\right)\, W^2 \,\phi (W)\right.\nonumber\\
& & +\, nt \left(\alpha(W) - W \alpha' (W)\right)^2 \left[
\right.\frac{[\alpha(W)
+\beta(W)]}{[\alpha(W) \beta(W)]^{1/2}} I_1
\left(2nt
[\alpha(W)\beta(W)]^{1/2}\right)
\nonumber\\
&&- \left. 2\left. I_0 
\left(2nt
[\alpha(W)\beta(W)]^{1/2}\right)
\right] \right\} \,\,,
\label{13}
\end{eqnarray}
where the functions $\alpha(W)$ and $\beta(W)$ have been defined in 
Eqs. (\ref{5}). 
It is readily verified that in the special case of a dichotomic 
velocity distribution
\begin{equation} \phi^{+}(V)=\phi^{-}(V)=\phi(V) = 
\frac{1}{2}[
\delta (V+c) + \delta (V-c)]\,,
\label{dichdistr}
\end{equation} 
one obtains from Eq. (\ref{13}) the exponential decay 
\begin{equation}
C(t)= c^{2}\,e^{- 2nct}\,\,.
\label{dich}
\end{equation} 
Similarly, the known result 
for a Maxwellian $\phi(V)$ is also recovered. 

$C(0)$ is of course $\int dW\, W^{2}\, \phi (W)$, and $C(t)$ initially 
decreases linearly with $t$, with a slope that works out to 
\begin{equation}
(dC/dt)_{t=0} = -2n \int_{-\infty}^{\infty} dW \,
\left\{W^{2} \phi(W) \alpha(W) + 
[\alpha(W) - W \alpha'(W)]^{2} \right\}\,\,.
\end{equation}
In particular, for a Maxwellian $\phi(V)$ one can evaluate 
the integral involved to obtain $(dC/dt)_{t=0} = -4n(k_{B}T/\pi 
m)^{1/2}$. In the general case, $C(t)$ is a non-monotonic function
of $t$, that becomes negative beyond a certain point and eventually
approaches zero from below as $t \rightarrow \infty$.
The long-time behavior of the velocity autocorrelation  
yields valuable information on the mixing properties 
and memory effects in
the system. The extraction of this behavior from Eq. (\ref{13})
is non-trivial. 
It is helpful  to note that $\alpha(W)$ is a non-increasing function of 
$W$, with $\alpha(W) \sim -W$ as $W \rightarrow -\infty$ and 
$\alpha(\infty) = 0$; whereas $\beta(W)$ is a 
non-decreasing function of $W$, with $\beta(-\infty) = 0$ 
and $\beta (W) \sim W$ as $W \rightarrow \infty$. Further,
whenever $\phi^{-}(V) = \phi^{+}(V) = \phi(V)$, we have
$\beta(W) = \alpha(-W) = W + \alpha (W)$.
An adequate number of terms in the asymptotic 
expansions of the Bessel functions and the other terms in the 
integrand in Eq. (\ref{13}) must be retained, consistent with the
fact that non-vanishing contributions to the integral come from the
region $W^{2}t \lesssim {\cal O}(1)$. As already mentioned, 
it has  been 
shown \cite{jep},\cite{lebo1} that $C(t)$ has a leading asymptotic behavior 
$\sim t^{-3}$ for a Maxwellian $\phi(V)$ \cite{criterion}. We have
corroborated this, and also extended the result to the next term in 
the asymptotic expansion: after a very lengthy 
calculation, we find
\begin{equation}
C(t) \sim
-\frac{1}{(nt)^{3}}\left(\frac{m}{2\pi 
k_{B}T}\right)^{1/2}\left(1-\frac{5}{2\pi}\right)\,
- \,\frac{1}{8(nt)^{4}}\left(\frac{m}{2\pi k_{B}T}\right) 
\left(177 - \frac{315 \pi}{16} 
-\frac{367}{\pi}\right) \,-\ldots\,\,.
\label{maxtail}
\end{equation}
Based on the emergence of a $t^{-3}$ tail in the Maxwellian case 
(and also for another extended distribution  $\phi (V)$ that falls off 
like $\arrowvert V \arrowvert ^{-3}$), as opposed to an 
exponential decay  (Eq. (\ref{dich})) for the 
dichotomic distribution, it has been concluded 
\cite{lebo1} that $C(t)$ decays exponentially if $\phi(V)$ is a
compact (``finite") distribution, and has a 
power-law tail whenever the support of $\phi(V)$ is  non-compact. 
However, as we now proceed to show, the actual criterion for the
emergence of a power law tail in $C(t)$ turns out to be {\it the
existence of a non-zero probability mass at $V = 0$ in 
$\phi(V)$}. (The physical reason for this will be described 
subsequently.) 
The issue is most clearly elucidated with the help of
two simple yet archetypical distributions $\phi (V)$ for which $C(t)$ 
can be determined analytically, and exact asymptotic expansions
obtained. These are, respectively, (i) a uniform distribution
in the finite interval $-c \leq V \leq +c$, and (ii) a discrete
distribution consisting of $\delta$-functions at $V = \pm c$ and
an additional one at $V = 0$. The former has a compact
support, and yet $C(t)$ turns out to have a $t^{-3}$ tail; while the latter, 
in contrast to the dichotomic distribution of Eq. (\ref{dichdistr}), 
leads in fact to an even heavier tail ($\sim t^{-3/2}$) for any
non-vanishing weight of the central $\delta$-function. We also
extend the result in Eq. (\ref{maxtail}) to the case of a 
general $\phi (V)$ that is sufficiently smooth at the origin.\\

\noindent (i) Accordingly, let us consider the uniform distribution
\begin{equation}
\phi(V) = \frac{1}{2c} \theta (V+c) \theta (c-V)\,\,. 
\label{19}
\end{equation}
For this distribution, $\alpha (W)$ is respectively equal to $- W$ 
for $W \leq -c$, and  
$(c-W)^{2}/4c$ for $-c \leq W \leq +c\,$; it vanishes identically for 
$W > c\,$. Recall also that $\beta (W) = W + \alpha (W)$. 
Inserting these in Eq. (\ref{13}) and carrying out the necessary 
calculations, we finally obtain a closed-form expression for $C(t)$.
In terms of the dimensionless time $\tau = nct$, this reads
\begin{eqnarray}
C(\tau) =
\frac{c^2}{4} \left[ \frac{1}{2} - \frac{1}{4\tau} + 
\left( \frac{1}{4\tau} +1 -
\tau \right) e^{-\tau}\,\, 
{}_{1}F_{1}\left(\frac{1}{2};\,\frac{3}{2};\,\tau\right) \right] \,\,,
\label{20}
\end{eqnarray} 
where ${}_{1}F_{1}$ is the usual confluent hypergeometric function. 
The slope at the origin is $(dC(\tau)/d\tau)_{\tau=0} = -\frac{2}{5} 
c^{2}$. The analytic form of $C(\tau)$ enables us to write down its 
exact asymptotic expansion for large $\tau$: 
\begin{equation}
C(\tau) \,\,\sim \,\, -c^{2} \sum \limits_{n=3}^{\infty} \frac{(n-1)(n-2)(2n-5)!!}
{\tau^n} \,\,.
\label{22}
\end{equation}
As mentioned earlier, this starts with an 
${\cal O}(\tau^{-3})$ term, although the support of $\phi(V)$ is compact.
Figure \ref{Cuniform} depicts the long-time behavior of the correlation
function.

\noindent (ii) Next, consider the discrete distribution 
\begin{equation}
\phi(V) = \mu \,\delta (V)+ \frac{1-\mu}{2} \,
[ \delta(V+c) + \delta (V-c)], \ 0 \leq \mu < 1 
\label{trich}
\end{equation}
which is an extension of the dichotomic distribution 
of Eq. (\ref{dichdistr}) to include an additional $\delta$-
function at $V = 0$ with a weight $\mu$.
Once again, $\alpha (W)$  vanishes identically for $W > c\,$. For
$W \leq c\,$, it is piecewise linear, being given by
\begin{equation}
\alpha (W) = \left\{
\begin{array} {ll}
    -W\,,& \,W < -c\\
    \frac{1}{2}(1-\mu)c - \frac{1}{2}(1+\mu)W\,, & \,-c \leq W < 0\\
    \frac{1}{2}(1-\mu)(c-W)\,,& 0 \leq W \leq c \,.
     \end{array}\right.
\label{trichalpha}
\end{equation}
We obtain in this case (with $\tau = nct$ as before)
\begin{equation}
C(\tau) = c^{2}\,(1-\mu) \,e^{-\tau}\left\{1 + 
(1-\mu)
\tau
\int \limits_0^1 du \,e^{\mu\,(1- u)\,\tau} 
 \left[
\displaystyle \frac{1-\mu +\mu u}{g(u)} I_1 \left(\tau g(u)\right) - 
I_0\left(\tau g(u)\right) \right] \right \}\,\,,
\label{24} 
\end{equation}
where $g(u) = \left( (1-\mu)(1-u)[(1-\mu)+(1+\mu)u] \right)^{1/2}\,$.
The slope at the origin is $(dC(\tau)/d\tau)_{\tau=0} = 
-(1-\mu)(2-\mu)c^{2}$. The long-time behavior in this case is
however quite different from that found in the previous cases.
Owing to the singularity in $\phi(V)$ at the origin, $\alpha'(W)$
has a jump at $W = 0$. As a consequence, the 
integrand in Eq. (\ref{13}) is now a function of 
$\arrowvert W \arrowvert$ rather than $W$. This leads to 
the occurrence of both even and odd powers of 
$\arrowvert W \arrowvert$ in the small-$W$  expansion
of the integrand, using which the asymptotic expansion of 
$C(t)$ is determined. For the latter, we now obtain
\begin{equation}
C(\tau) \sim  -\displaystyle \frac{\mu\,c^{2}}{ \tau^{3/2}}\,\,
\frac{(1-\mu)^{1/2}}{4\,
\,(2\pi)^{1/2}} \, 
-\displaystyle \frac{\mu\,c^{2}}{\tau ^{5/2}}\,\,\frac{3  (3-2\mu^{2})}
{32\,\,(2\,\pi)^{1/2}\,(1-\mu)^{1/2}}\,+\ldots\,\,.
\label{Cmu}
\end{equation}
Thus $C(t)$ now has an even {\it slower} power-law 
decay, starting with an ${\cal O}(t^{-3/2})$ term, 
as long as $\mu \neq 0$, i.e., as long as there is 
a finite probability mass at $V=0$. When $\mu =0$, all the terms in 
the asymptotic expansion vanish, and $C(t)$ reverts to  
the exponential decay that obtains in the case of the dichotomic distribution, 
Eq. (\ref{dich}). Figure \ref{Ctrich} shows the long-time behavior
of the correlation function for different values of the weight 
parameter $\mu$, including (for ready comparison) the case $\mu =0$.

The relative roles of the $\delta$-functions in $\phi(V)$
at $V=0$ and at $V= \pm c$ may be examined a little more closely. This 
aspect is not so transparent in the representation of Eq. (\ref{24}) for
$C(t)$, but is made
more manifest with the help of its Laplace
transform. This enables us to write $C(t)$ in the 
form 
\begin{eqnarray}
\hspace{-0.5cm}C(t) &=&(1-\mu) c^{2}\left\{
\frac{e^{-\frac{2nct}{1+\mu}}}{1+\mu}\right.\nonumber\\
&+& \left.\mu \,{\cal{L}}^{-1}
\frac{\left (s \mu + [s^{2} + 2snc(1-\mu)]^{1/2}\right)}
{\left((1+\mu)s + 
2nc\right)\left(s + \mu [s^{2} + 2snc(1-\mu)]^{1/2}  
+ 2nc(1-\mu) \right)}\right\}\,\,.
\label{Ctilde}
\end{eqnarray}
where ${\cal{L}}^{-1}$ denotes the inverse Laplace transform. 
Comparing this with the pure exponential decay $c^{2}\,e^{- 2nct}$
that obtains for the dichotomic velocity distribution, we see that
the $\delta$-function in $\phi(V)$ at $V=0$ is entirely responsible
for the second term (which vanishes when $\mu = 0$). Further,
the timescale in the exponential part is itself modified
from the usual correlation time for a dichotomic process, which is
$(2nc)^{-1}$ in the present context, 
to $(1+\mu) (2nc)^{-1}$, as one might expect on physical
grounds.

We are now in a position to understand the physical origin of the 
power-law tail in $C(t)$. The particles of the 
system do not undergo any systematic drift in the homogeneous
case. Going back to an inspection of the manner in 
which the particle under consideration skips from one free trajectory
to another through  
collisions, we see that, if the stationary velocity 
distribution $\phi(V)$ of the gas particles has {\it a finite 
probability mass at} $V = 0$, the particle will repeatedly
find itself on a 
trajectory with zero slope, i. e., revert to the zero (= average)
velocity state. This persistence is like a
memory effect, and it shows up as a 
slow (power-law) decay of $C(t)$. The compactness or otherwise of
the support of $\phi(V)$ does not play a role as far as
this aspect is concerned.

It is also possible to find the precise conditions
under which  the leading asymptotic behavior of $C(t)$ starts
with a $t^{-3}$ term: this is so if $\phi(V)$ is at 
least twice differentiable at
the origin, and moreover $\phi(0)$ and  $\phi''(0)$ do not
both happen to be zero. (We recall that $\phi(V)$ has been taken to be 
a symmetric function, so that all its derivatives of odd order 
vanish at the origin.) The general asymptotic expansion of 
$C(t)$, for a distribution $\phi(V)$ which is differentiable a sufficient 
number of times at $V = 0$, reads:
\begin{eqnarray}
C(t) &\sim &
-\frac{1}{(nt)^{3}}
\left[\phi(0)-6\,\alpha(0)\phi^{2}(0)-\alpha^{2}(0)\phi''(0)\right]\,\nonumber\\
&-&\frac{1}{256\, \alpha(0)\,(nt)^{4}}\left[-315\,\phi(0)\,
+3456\, \alpha(0)\phi^{2}(0)-2880 \,
\alpha^{2}(0)\phi^{3}(0)\right.\nonumber\\
&+&\left.19840\,\alpha^{3}(0)\phi(0)\phi''(0)
-2208\, \alpha^{2}(0)\phi''(0) -256\, 
\alpha(0)\phi^{(iv)}(0)\right]\,\, 
\,-\ldots\,\,.\nonumber\\
\label{asim}
\end{eqnarray} 
This extends the result presented in Eq. (\ref{maxtail}) for a 
Maxwellian. Thus, for a distribution $\phi(V)$
that is {\it regular} at the 
origin and has a non-vanishing derivative of some finite order at that 
point, implying that there is a non-zero probability mass at
$V = 0$, $C(t)$ will certainly have a power-law decay:  If 
$\phi(0) \neq 0$, the leading term is generically $\sim t^{-3}\,$; on 
the other hand, if $\phi(0) = 0$ and its first non-vanishing derivative 
at the origin is its $(2r)$-th derivative, the leading term in $C(t)$ 
is $\sim t^{-r-2}$. 

\section{The inhomogeneous system}

We turn now to the inhomogeneous system, in which the particles to 
the left and right of the piston are initially in different macroscopic states
specified by $(n^-,\phi^-)$ and $(n^+,\phi^+)$ respectively. The 
piston now has, in general, a non-vanishing mean drift
velocity that asymptotically approaches $\overline{W}$. However,
as we shall see, the variance of its position indeed increases linearly 
with time. The quantity of interest is therefore the effective diffusion 
coefficient $D$, which we shall determine. We also find the leading
asymptotic behavior of the other moments of the position (about its
mean value). 

In the homogeneous system, $D$
is of course equal to ${\int}_{0}^{\infty} \,C(t) \,dt$ (this
integral being  absolutely convergent for the system at hand).
However, owing to the non-stationarity of the velocity 
autocorrelation in the inhomogeneous case, $D$ must now be computed 
directly from 
the long-time behavior of the mean square displacement of the piston. 
The asymptotic behavior of the piston does not depend on its initial 
state. We can therefore set $V_{0}=0$ in $P(X,V,t \arrowvert 
0,V_{0},\,0)$ (Eq. (\ref{4})) and integrate it over $V$ to
calculate
the position
distribution function $p(X,t)$ of the piston. Using the fact that
the derivatives of the functions $\alpha$ and $\beta$ are given by
\begin{equation}
\alpha' (W) = - \int \limits_W^\infty dU \phi^- (U)\,\,\,\,\ \mbox{and}\ 
\,\,\,\,\beta' (W)
= \int\limits_{-\infty}^W dU \phi^+ (U)\,,
\label{27}
\end{equation}
we find
\begin{eqnarray}
p(X,t)& =& e^{-t\left[n^- \alpha(0) +
n^+ \beta(0)\right]} I_0 
\left(2t
[n^- \alpha (0) n^+ \beta (0)]^{1/2} \right)
\,\delta(X)\,
\nonumber\\ 
& +&\,e^{-t\left[n^-\alpha ({X}/{t}) + n^+ 
\beta({X}/{t})\right]}
\left\{ \left[n^- \theta (X) |
\alpha' (X/t) | + n^+ \theta (-X) \beta'(X/t)\right]\right. 
\nonumber\\ 
& \times &I_0 
\left(2t
[n^- \alpha (X/t) n^+ \beta (X/t)]^{1/2} \right)
+ \left[n^-\theta(-X)|\alpha'(X/t)| 
\left(n^+ \beta (X/t)/n^-\alpha (X/t)\right)^{1/2}
\right. \nonumber\\
&+&\left.\left. n^+ \theta(X) \beta' (X/t)
\left(n^- \alpha (X/t)/n^+\beta (X/t)\right)^{1/2}
\,\,\right]\,I_1 
\left(2t
[n^- \alpha (X/t) n^+ \beta (X/t)]^{1/2} \right)
\right\} \,\,.\nonumber\\
\label{28}
\end{eqnarray}
The variance of the position is given by
\begin{equation}
\int_{-\infty}^{\infty}\,dX \,\left(X-\langle X(t)\rangle\right)^2\,p(X,t) \,\,.
\label{29}
\end{equation}
In the long-time limit, $\langle X(t) \rangle  = \overline{W}\,t$.
Using the asymptotic behavior of the Bessel functions in $p(X,\,t)$,
the leading behavior of the variance is given by
\begin{eqnarray}
\langle (X-\overline{W}\,t)^{2} \rangle &\sim&\frac{t^{5/2}}
{2(\pi n^{-} \alpha(\overline{W}))^{1/2}} \left[n^{-} \arrowvert 
\alpha'(\overline{W})\arrowvert + n^{+}\beta'(\overline{W})\right] 
\int_{-\infty} ^{\infty} dW (W-\overline{W})^{2} \nonumber\\
&&\hspace{-1cm} \times \exp\left\{-t\,(W-\overline{W})^{2} \left([n^{-}/4 
\alpha(\overline{W})]^{1/2} \,\arrowvert \alpha'(\overline{W}) \arrowvert +
[n^{+}/4 
\beta(\overline{W})]^{1/2} \, \beta'(\overline{W}) \right)^{2} 
\right\} 
\label{variance}
\end{eqnarray}
which simplifies to $2Dt$, with a diffusion coefficient given by
\begin{equation}
D = \frac{n^-\alpha (\overline{W})}{\left[n^-|\alpha'(\overline{W})| + n^+\beta'
(\overline{W})\right]^2}  = \frac{1}{2} \frac{n^-\alpha (\overline{W}) +
n^+ \beta (\overline{W})}{\left[n^-|\alpha' (\overline{W})| + n^+
\beta'(\overline{W})\right]^2}\,\,.
\label{30}
\end{equation} 
This is the general formula sought. 

We first note (as a check) that in the special case of the homogeneous system,
Eq. (\ref{30}) reduces to 
\begin{equation}
D = \,\,\frac{\alpha(0)}{n} \,= \,\frac{1}{n} \int \limits_0^\infty 
dU\, U \,\phi (U)
\,=\,
\frac{\langle|U|\rangle}{2n} \,\,,
\label{31}
\end{equation}
in agreement with the known result \cite{lebo1}.
As mentioned earlier, in this case $D$ must also be equal to
the integral of the velocity  autocorrelation $C(t)$. We have verified 
that this is indeed so.

Some interesting special 
cases emerge from the general formula of Eq. (\ref{30}). 
If the densities $n^-$ and  $n^+$ are such that the 
drift velocity $\overline{W} = 0$ even though one has  
different (but symmetric) distributions $\phi^{+}(V)$ and 
$\phi^{-}(V)$ on either side of the piston, the formula for $D$ 
simplifies somewhat. Since 
$\alpha'(0) = -\beta'(0) = -{1}/{2}$, we find
\begin{equation}
D = \frac{n^- \alpha(0)}{(n^- + n^+)^2}\,. 
\label{33}
\end{equation}
In particular, if $\phi^\pm$ are Maxwellians (with $n^-\sqrt{T^-} = n^+\sqrt
{T^+}$ to ensure that $\overline{W} = 0$),
\begin{equation}
D = \frac{n^-}{(n^-+n^+)^2} \left( \frac{8k_{B}T^-}{\pi m} 
\right)^{1/2}\,\,.
\end{equation}

On the other hand, if 
$\phi^{-}(V)  = \phi^{+}(V) = \phi(V)$, but the system remains 
inhomogeneous because 
$n^{+} \neq n^{-}$, then $\overline{W} \neq 0$. For the compact 
uniform velocity
distribution of Eq. (\ref{19}), we find 
\begin{equation}
\overline{W} =  c \,\,\frac{\sqrt{n^{-}} - \sqrt{n^{+}}}{\sqrt{n^{-}} + 
\sqrt{n^{+}}}
\label{Wbaruniform}
\end{equation}
and
\begin{equation}
D = \frac{c}{(\sqrt{n^-} + \sqrt{n^+})^2}\,.
\label{Duniform}
\end{equation}
For the discrete distribution of Eq. 
(\ref{trich}), we have 
\begin{equation}
\overline{W} =  c \,\,\frac{(1-\mu)(n^{-} - n^{+})}{n_{>}(1-\mu) + 
n_{<}(1+\mu)}\,, 
\end{equation}
where  $n_{>} = \mbox{max}(n^{-}\,,\,n^{+})\,$,   
$\,n_{<} = \mbox{min}(n^{-}\,,\,n^{+})$. The corresponding diffusion
coefficient is found to be
\begin{equation}
D = \frac{4c\,n^{-} n^{+}(1-\mu)}{[n_{>}(1-\mu) + 
n_{<}(1+\mu)]^{3}}\,.
\label{Dtrich}
\end{equation}
It is noteworthy that the interplay of the central $\delta$-function in 
the velocity distribution and  the inhomogeneity due
to the different densities on either side of the piston affects even
the {\it diffusion} coefficient.
Setting $\mu = 0$ in the above yields the corresponding expressions for 
the dichotomic distribution of Eq. (\ref{dichdistr}).

The leading asymptotic behavior of the higher moments 
$\langle (X-\overline{W}\,t)^{r} \rangle$ can also be determined. 
Here we merely quote the salient result obtained.
For the even moments $r = 2l$, it is straightforward to show (along 
the same lines as in the case of the variance) that
$\langle (X-\overline{W}\,t)^{2l} \rangle \sim {\cal O}(t^{l})$.
The calculation is more involved for the odd moments $r = 2l + 1$,
but the final result is that 
$\langle (X-\overline{W}\,t)^{2l+ 1} \rangle \sim {\cal O}(t^{l})$
as well. As the expressions obtained for the precise
coefficients are lengthy, we do not write them down here.

\section{Interpretation as a stochastic process}
We conclude by showing that (in the thermodynamic limit) the form of 
the 
distribution of the position of the piston, in fact that of {\it any}
of the particles in the system, is effectively that of a stochastic 
process driven
by a noise that can be given a direct physical 
interpretation. 

Conventionally, the stochastic approach to single-particle 
dynamics in a many-body system  begins with its modeling by 
a stochastic evolution equation involving noise terms
with prescribed statistical properties. 
One then extracts the 
corresponding properties of the driven variable(s). Here, however, 
we have the converse situation. The exact 
time-dependent one-particle distributions
are known,
and the task is to identify 
the stochastic process to which the complicated dynamics effectively 
reduces, at least as far as the one-particle dynamics is concerned. 
What kind of stochastic process does
the position $X(t)$ of the piston (or any other particle) represent,
after the averaging over the initial states of the gas particles is done, 
and the thermodynamic limit 
taken? And in what kind of ``noise" are the combined effects of the 
other particles 
in the system encapsulated? 

It is evident from the rather complicated
expressions for $P(X,V,t|0,V_0,\,0)$ and the reduced distributions derived 
from it that $X(t)$ is 
unlikely to satisfy any simple or standard stochastic differential 
equation; nor does $p(X,t)$
appear to be the solution of any simple master equation - in 
particular, of any obvious partial differential equation of 
finite order. Intricate correlations exist, that cannot be neglected.
The effects of recollisions are obviously significant, a direct
instance being provided by the form of the first term in Eq. (\ref{4}).
This term 
represents the probability for the piston to find itself in its initial 
state at time $t$. Now,  
the probability that the initial state of the piston 
($X = X_{0}=0$, $V = V_0$) {\it persists} till time $t$ (i.e., the 
piston suffers no collisions till time $t$) is easily shown to 
be simply  $\exp \,[-t(n^-\alpha (V_0) + n^+ \beta
(V_0))]$. Thus the extra factor 
$I_0$ in  the term proportional to $\delta (X-V_0t) 
\,\delta(V-V_0)$ in $P(X,V,t|0,V_0,\,0)$ is entirely due to the 
effects of recollisions \cite{lebo1}. As 
the concept of an effective noise is only meaningful in the thermodynamic 
limit and 
when ergodicity obtains, we must examine for this purpose the 
structure of the terms in the solutions
{\it other} than the ones arising from the 
returns to any specific initial state. 

The occurrence of the Bessel functions
$I_0$ and $I_1$ in Eqs. (\ref{4}) and (\ref{28}) seems to suggest 
some sort of link with dichotomic 
diffusion (i.e., the integral of a dichotomic Markov process) and
the well known telegrapher's equation and its solution. Indeed, 
in the homogeneous case, with $\phi (V)$ equal 
to the dichotomic velocity distribution of Eq. (\ref{dichdistr}) and
$V_{0} = \,\pm \,c\,$, Eq. (\ref{28}) for $p(X,t)$ does reduce  
to the solution correponding to dichotomic diffusion \cite{bala}, 
once again except for the extra factor of 
$I_0 (nct)$ in the ``ballistic" term representing the probability of
the occurrence of the initial state at time $t$. But this
does not explain the origin of the Bessel functions in the general
case. Nor does it really do so even in the special case referred to,
other than the not-very-helpful observation that the solution of the
telegrapher's equation involves $I_0$ and $I_1$. As the effective ``noise" 
we seek should be essentially the same for every particle, our arguments
should indeed apply to any of the particles, and not just the 
piston. Proceeding 
as in the case of the piston (i.e., averaging over the initial 
positions and velocities of all the particles except the piston, 
and with $X_{0} = 0, V_{0} = 0$), we find the following result for the
position distribution of the $b\,$th particle at time $t$:
\begin{eqnarray}
p_{b}(X,t)& =& e^{-t\left[n^- \alpha(0) +
n^+ \beta(0)\right]} 
\left(n^+ \beta (0)/n^-\alpha (0)\right)^{b/2}
I_{b} 
\left(2t
[n^- \alpha (0) n^+ \beta (0)]^{1/2} \right)
\,\delta(X)\,\nonumber\\ 
& +&\,e^{-t\left[n^-\alpha ({X}/{t}) + n^+ 
\beta({X}/{t})\right]}
\left(n^+ \beta (X/t)/n^-\alpha (X/t)\right)^{b/2}\nonumber \\
&\times &\left\{ \left(n^- \theta (X) |
\alpha' (X/t) |\right.\right.
+\left. n^+ \theta (-X) \beta'(X/t)\right) 
I_b \left(2t [n^- \alpha (X/t) n^+ \beta (X/t)]^{1/2} 
\right)\nonumber \\
&+& n^-\theta(-X)|\alpha'(X/t)| 
\left(n^+ \beta (X/t)/n^-\alpha (X/t)\right)^{1/2}
I_{b+1} 
\left(2t
[n^- \alpha (X/t) n^+ \beta (X/t)]^{1/2} \right) \nonumber\\
&+& \left. n^+ \theta(X) \beta' (X/t)
\left(n^- \alpha (X/t)/n^+\beta (X/t)\right)^{1/2}
\,\,\,I_{b-1} 
\left(2t
[n^- \alpha (X/t) n^+ \beta (X/t)]^{1/2} \right)
\right\} \,\,.\nonumber\\
\label{bthparticle}
\end{eqnarray}
Here $b \in\mathbb{Z}$. (Setting $b = 0$ one recovers the result in
Eq. (\ref{28}) for the piston, remembering that $I_{-1} = I_{1}$.) 
The occurrence of the Bessel functions $I_{b+1}$, $I_{b}$ and $I_{b-1}$ 
shows quite clearly that we must 
look for a link to Bessel functions other than one via the
solution to the telegrapher's equation. 

The example of the dichotomic velocity distribution 
in the homogeneous case does
provide a valuable clue, though. Let us therefore examine this case 
for a moment, focusing on the piston, and taking $V_{0}$ also to 
be either $c$ or $-c\,$, to mask the effects of any special initial 
conditions. The actual ``free" trajectories of all the  
particles are then straight lines with slopes
restricted to the values $\pm \,c$. A little thought shows that after 
its first collision, 
the piston alternately `rides' on a free trajectory belonging to the 
gas on its right, and one belonging to the gas on its left. For 
brevity,
we shall refer to these as `right' and `left' trajectories. (Which of 
these the piston gets on to first, depends on whether its initial 
velocity $V_{0}$ is equal to $c$ or $-c$.) Moreover, the piston is 
alternately hit by a `right' particle with velocity $-c$ and a `left' 
particle with velocity $+c$. The number of `right' collisions minus 
the number of `left' collisions can only take on the values $+1$, $0$ 
and $-1$.  The resulting 
zig-zag path is precisely that of a particle whose position $X$ satifies 
the stochastic
differential equation $\dot{X}= c \,\xi(t)$, where $\xi(t)$ is a
stationary dichotomic Markov process (DMP) alternating between the values
$\pm 1$ with a certain mean switching rate $\lambda$. Such a DMP is 
generated by  
a stationary Poisson pulse process of intensity 
$\lambda$. One could also regard it as made up of two independent
Poisson pulse processes, each with an intensity $\lambda/2$,
alternating with each other. This would seem to be a little 
more closely linked to the present
situation, where one might imagine the two states of $\xi(t)$ to be
related in some sense to the piston being on a right trajectory 
and a left trajectory, respectively. But 
the connection is still far from obvious, and requires some more work.  

Let  $\nu^{+}$ and $\nu^{-}$ be  two independent stationary Poisson 
processes with respective intensities (i.e., mean rates) 
$\lambda^{+}$ and $\lambda^{-}$, 
so that their mean values at time $t$ are  $\lambda^{+}\,t$ and 
$\lambda^{-}\,t$. 
It is easily shown that their {\it difference} $(\nu^{+} 
-\nu^{-})$, which can take on any integer value, 
has a time-dependent distribution given by 
\begin{equation}
\mbox{Pr}\, (\nu^{+} -\nu^{-} = r;\,t) = e^{-(\lambda^{+} + \lambda^{-})\,t}
\,\,(\lambda^{+}\,t/ \lambda^{-}\,t)^{r/2}\,I_{r}
\left(2\sqrt{(\lambda^{+}\,t)\,(\lambda^{-}\,t)}\,\right)\,,\,\, r \in
\mathbb{Z}\,\,.
\label{poissondiff}
\end{equation}
It is {\it this} distribution that holds the key to understanding
the structure of the one-particle distributions in the problem under
consideration.
Let the piston be at a position $X > 0$ at time $t$, on a (segment)
of a free trajectory. Translating the entire system to bring the initial 
coordinate of this trajectory to the origin, the
instantaneous velocity of the piston is $X/t$ (recall that the 
trajctories are all straight lines, a direct consequence of the equal 
mass condition). It can be hit by a 
left particle provided the latter has a positive 
velocity ($+\,c\,$ in the case under consideration) that is greater 
than $X/t$. The mean rate at which 
this happens is given by the product of the concentration $n$ of the
gas, the magnitude of the {\it relative} velocity $(c - X/t)$, and the
probability 
$\frac{1}{2}$ that
the velocity of the gas particle is $c$ (see Eq. (\ref{dichdistr})):
in other words, $\lambda^{-} = \frac{1}{2}\,n(c - X/t)$. Similarly, 
the mean rate at which the piston is hit by a right particle is given 
by $\lambda^{+} = \frac{1}{2}\,n(c + X/t)$. Moreover,
the number of `right' collisions minus 
the number of `left' collisions only takes on the values $+1$, $0$ 
and $-1$. Putting in all the foregoing facts and 
their obvious extension to the 
case $X < 0$, and transforming from the random variable 
$(\nu^{+} -\nu^{-})$ to $X$,
we are led to the expression 
\begin{eqnarray}
\frac{n}{2} 
e^{-nct}\left\{
I_{0}\left(n(c^{2}t^{2}-X^{2})^{1/2}\right) 
+\,\left((ct+X)/(ct-X)\right)^{1/2} 
I_{1}\left(n(c^{2}t^{2}-X^{2})^{1/2}\right)\right\}\nonumber\\
\times \,\theta(X+ct)\,\theta(ct-X)\,.
\label{pxtdichot}
\end{eqnarray}
This is precisely the solution for $p(X,t)$
to which  Eq. (\ref{28}) reduces in this special case, {\it apart} from 
the contribution from the initial state. The latter is 
$e^{-nct} \,\delta (X \pm \, ct)$ when $V_{0} = \pm \,c$, and 
$e^{-nct} 
I_0\left(n ct \right) \,\delta (X)$ when $V_{0} = 0$. 
It is also worth noting how the
factors of $2$ (coming from the formula of Eq. (\ref{poissondiff})) and 
$\frac{1}{2}$ (coming from the rates $\lambda^{\pm}$) cancel out in the 
argument of the Bessel functions in (\ref{pxtdichot}).

These arguments are extended to the general inhomogeneous case as
follows. 
When the piston is at position $X$ at time $t$, collision with a 
left particle is possible
provided the latter has a velocity $U$ in the range $(X/t, \,\infty)$, 
and the magnitude of the relative velocity is $(U - X/t)$. Since the
gas on the left has a concentration $n^{-}$, and the velocities of its 
particles are drawn from the distribution $\phi^{-}$, the effective 
mean rate of `left' collisions of the piston is given by 
\begin{equation}
\lambda^{-} = n^{-} \,\int\limits_{X/t}^{\infty} dU \phi^{-} (U) 
(U-X/t) \,,
\label{lambdaminus}    
\end{equation}
which is nothing but $n^{-}\alpha (X/t)\,$. Similarly, in the same given 
state the piston
can only be hit by a right particle with a velocity in the range 
$(-\infty,\, X/t)$, and it follows that 
\begin{equation}
\lambda^{+} = n^{+} \,\int\limits_{-\infty}^{X/t} dU \phi^{+} (U) 
(X/t - U) \,= \, n^{+}\beta (X/t)\,,
\label{lambdaplus}    
\end{equation}
since it is the {\it magnitude} of the relative velocity that appears 
in the mean collision rate. This explains the genesis and form of the 
argument of the Bessel functions in Eq. (\ref{28}). 
The extra factors involving  $n^{-}|\alpha' (X/t)|$ and 
$n^{+}\beta'(X/t)$ that appear in the expression for $p(X,t)$ in 
Eq. (\ref{28}) are just the Jacobians that arise when we 
transform from the distribution of $(\nu^{+} -\nu^{-})$ to that 
of $X$. Finally, although the piston 
{\it can} have successive left collisions or right collisions in the 
general case, unlike what happens in the case of
the dichotomic velocity distribution, the remarkable fact is 
that their contribution to the probability distributions seems to 
vanish in the system at hand. The number of `right'
collisions minus 
the number of `left' collisions only takes on the values $+1$, $0$ 
and $-1$ even in the general case,
presumably as a consequence of the smearing 
out implied by the averaging and the 
thermodynamic limit. 
Likewise, for the $b\,$th particle from the piston, the difference
between the number of times the particle has occupied a right trajectory 
and the number of times it has occupied a left trajectory up to 
any time $t$ 
predominantly takes on the values
$b$ and $b \pm 1$ in this limit. The central value $b$ of this 
difference is in fact 
a consequence of its initial ordering, that is not altered by 
the collisions. The foregoing provides an understanding of the 
structure of the 
terms (other than the contribution due to the initial state)
in Eq. (\ref{bthparticle}). 

In conclusion, we see that the motion of any particle in the system 
may be regarded, in the 
thermodynamic limit, as being driven by two independent Poisson pulse 
processes representing the effects of the gases on the left and right of the 
central particle. The intensities (mean rates) 
of these processes
have the direct physical interpretation given above with regard to 
Eqs. (\ref{lambdaminus}) and (\ref{lambdaplus}). As a Poisson
process {\it per se} is an uncorrelated pulse process, 
each particle is effectively subjected to two independent noises in 
this precise sense. However, as the intensity of each noise is 
state-dependent (the driven variable $X$ appears explicitly in the 
limits of integration in $\alpha (X/t)$ and $\beta (X/t)$), the 
flow of $X$ is not given by any simple stochastic differential equation 
with additive or even multiplicative noise, which is only to be 
expected.

\acknowledgments {This research is supported by the 
Interuniversity Attraction Poles program of the Belgian Federal 
Government and the Belgian National Fund for Scientific Research.
VB acknowledges the warm hospitality of the Limburgs 
Universitair Centrum and the Universit\'e Libre de Bruxelles.}

\newpage

\begin{figure}[h]
\begin{center}
\label{Cuniform}
\caption{ Asymptotic behavior of the normalized velocity autocorrelation 
$C(t)/C(0)$  as a function
of time (in units of $1/nc$), for
a uniform distribution $\phi (V)$ (Eq. (\ref{19})).}
\end{center}
\end{figure}

\begin{figure}[h]
\begin{center}
\label{Ctrich}
\caption{ Asymptotic behavior of the normalized velocity autocorrelation 
$C(t)/C(0)$ as a function
of time (in units of $1/nc$) for different values of $\mu$,
for the discrete
distribution $\phi (V)$ in Eq. (\ref{trich}).}
\end{center}
\end{figure}


\begin{thebibliography}{99}
\bibitem{criterion} As one invokes 
the asymptotic behavior of 
$I_{n}$ for large argument in Eq. (\ref{13}) to obtain the
long-time behavior of $C(t)$, the onset of the latter is given
by the condition
$t >> (\pi m/2n^{2}k_{B}T)^{1/2}$ for a Maxwellian $\phi (V)$.
\end{thebibliography}
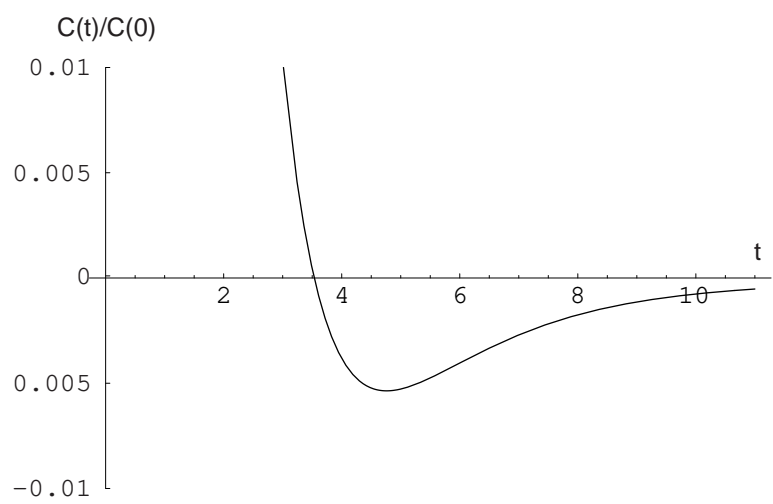
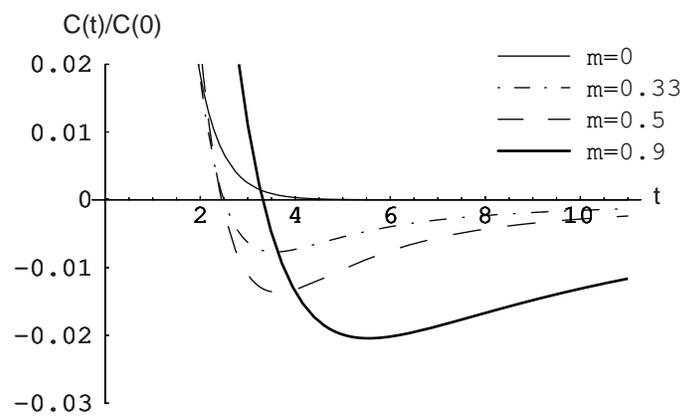
\end{document}